\documentclass[fleqn,12pt,twoside]{article}
\usepackage[headings]{espcrc1}

\usepackage{graphicx}
\usepackage[figuresright]{rotating}

\newcommand{\AmS}{{\protect\the\textfont2
  A\kern-.1667em\lower.5ex\hbox{M}\kern-.125emS}}

\title{Deep sub-barrier fusion reactions of the light nuclei $^{12}$C and 
$^{16}$O}

\author{\c S. Mi\c sicu\address[NIPNE]{NIPNE-HH, Department for 
Theoretical Physics,\\
Bucharest-Magurele, POB MG-6, ROMANIA }%
 \thanks{
This work was supported by
CNCSIS Romania, under Programme PN-II-PCE-2007-1, Contract No.49 (\c S.M. and 
F.C.). }
and
F. Carstoiu\addressmark[NIPNE]}

\runtitle{Deep sub-barrier fusion reactions  of the light nuclei $^{12}${\rm C} 
and
$^{16}${\rm O}}
\runauthor{\c S. Mi\c sicu}

\begin{document}

\maketitle

\begin{abstract}
\small Fusion reactions relevant for the carbon and oxygen burning cycles
in highly evolved stars are investigated in a standard approach to fusion.
We employ the double folding method to evaluate the ion-ion interactions with 
Gogny-D1 and M3Y-Paris $n-n$ effective forces. The cross-section evaluation do not
indicate a possible hindrance even at the lowest  energies under the barrier.
Reaction rates at temperatures relevant for the  stellar
processes are estimated and compared to the traditional and modern
extrapolation  formulas.
\end{abstract}

\vskip 0.5cm

The experimental data on sub-barrier fusion, accumulated in last
years, disclosed a new phenomenon, which consists in a severe
reduction of the cross sections once the bombarding energy 
is approaching a threshold value \cite{jiang02}. Until recently 
the application of the standard coupled channel model failed in
resolving the cross-section hindrance puzzle. Starting with 2005 
we performed a systematic investigation of the various fusion 
reactions exhibiting this phenomenon at low bombarding energies.
The fundamental ingredient in our approach was the nucleus-nucleus
potential, which differs from the traditional one, such as the 
Woods-Saxon or Aky\"uz-Winther, by an additional term dictated 
by the necessity that nuclear matter saturates. The result of this
modification is a massive change of the nucleus-nucleus potential 
inside the Coulomb barrier. From quantum-mechanical point of view 
we simply deal with a decrease of the transmission probability across
the barrier and consequently a hindrance in fusion.  
We succeeded in obtaining a very good description of the fusion cross-sections
for medium-heavy  projectile target combinations 
\cite{misesb06a,misesb06b}, medium-heavy  projectile and medium-light 
target \cite{jiang06}, medium-light projectile and target \cite{jiang08}, 
light projectile and heavy target \cite{esb07}. 
Typical to all the reactions displaying fusion hindrance at energies deep 
under the barrier is also the fact that the $S$-factor has an apparent maximum 
near the threshold energy. 

Since the hindrance phenomenon occurs over a wide range of heavy-ion masses,
it is then of interest to search also for other projectile-target combinations      
that might display the same hindrance features.
Very recently, it was conjectured by Jiang et collab. that the hindrance
could also affect the stellar reactions rates \cite{jiang07}

\begin{figure}[t]
\center{
\includegraphics[scale=0.52]{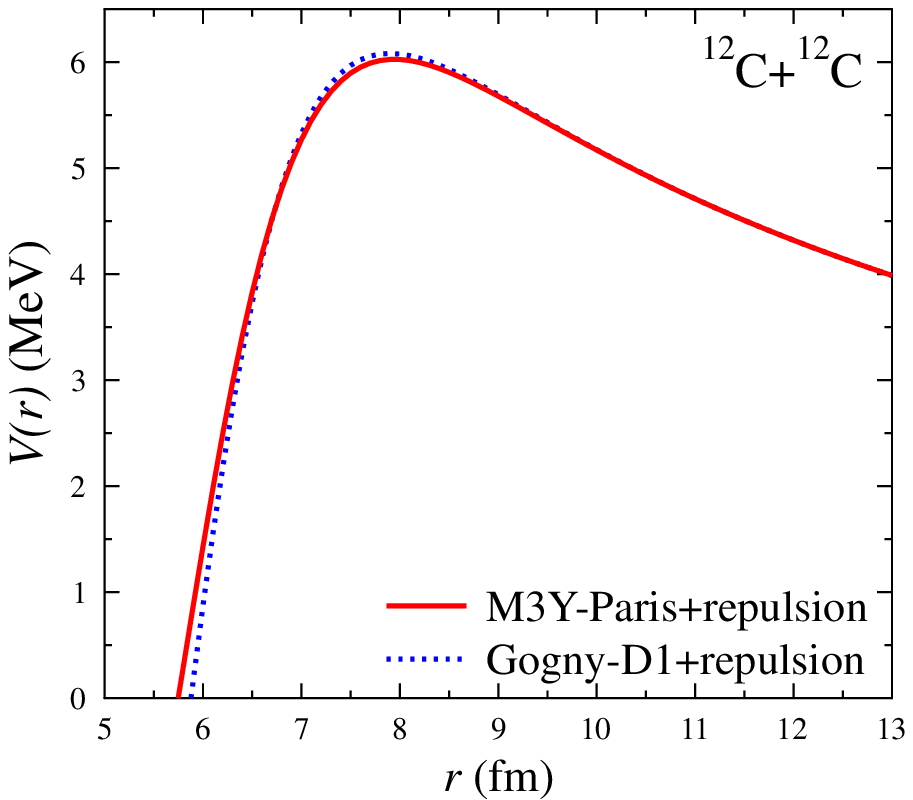}
\includegraphics[scale=0.52]{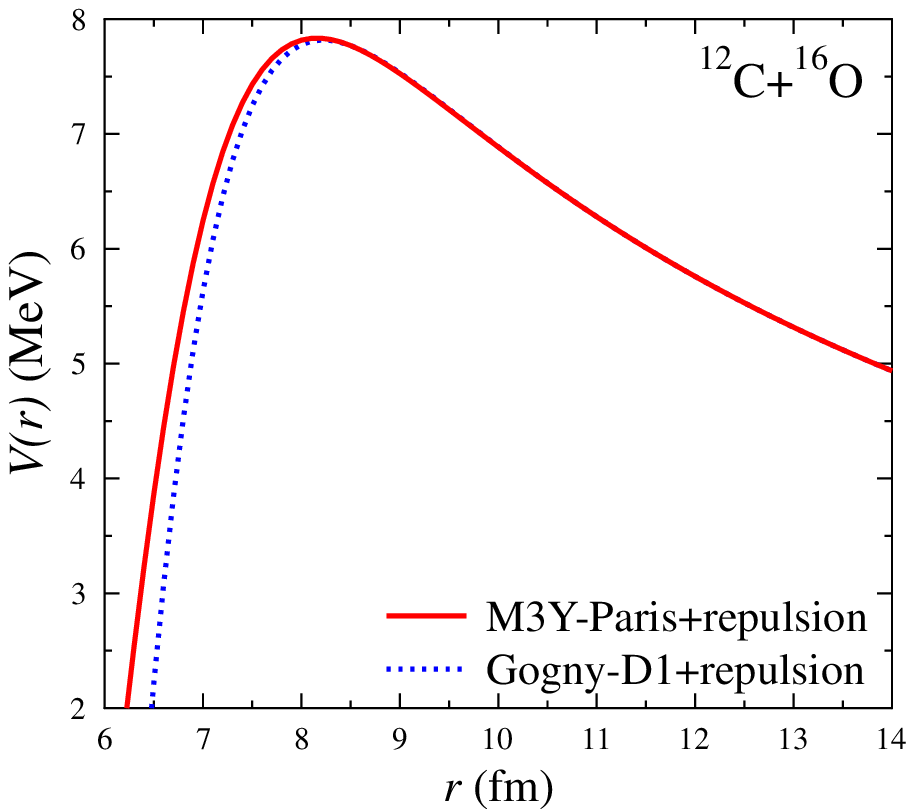}
\includegraphics[scale=0.52]{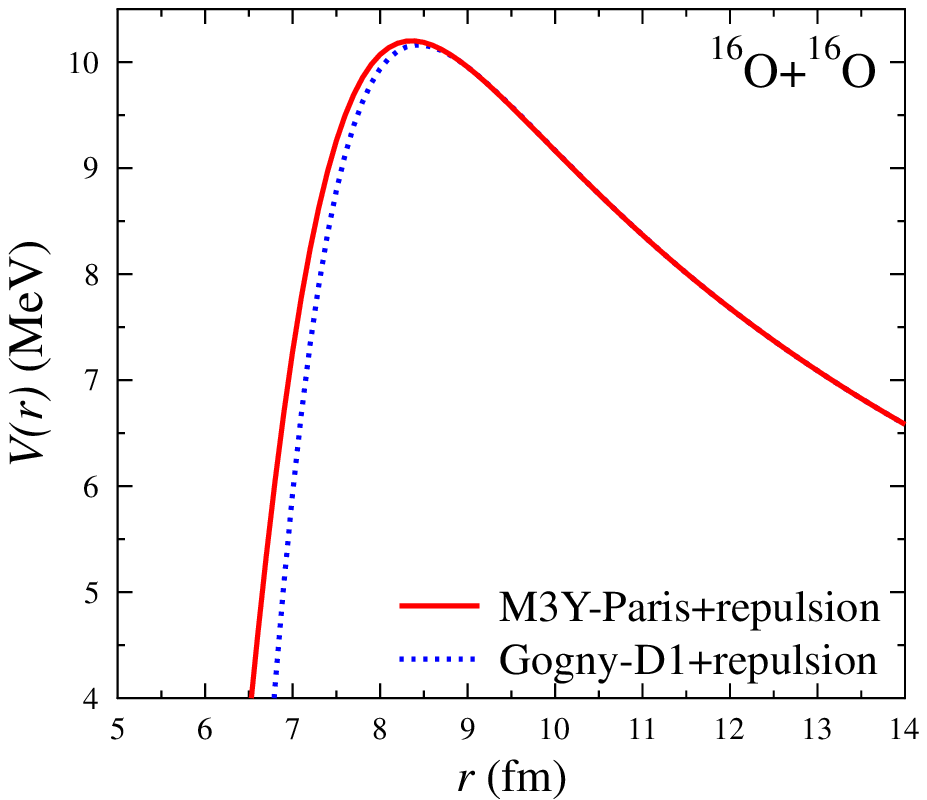}}
\caption{Ion-ion potentials for the three reactions investigated in this
work. Solid curve is the potential based on the M3Y-Paris $n-n$ effective
interaction whereas the dotted line was obtained with the Gogny-D1.}
\label{Fig:1}
\end{figure}

Although the heavy-ion literature is abundant in studies
on the sub-barrier reactions involving $^{12}$C and $^{16}$O we reinvestigate
in this work their fusion cross-sections using the double-folding heavy-ion potentials derived from finite-range effective $n-n$ forces, with or without density-dependent 
contributions. We provide a more accurate local equivalent of the non-local exchange 
potential by using the Perrey-Saxon procedure and the densities 
of the reacting nuclei were derived from a spherical Hartree-Fock calculation 
using the density functional of Beiner and Lombard \cite{beilom74}. 
The strength of the surface term in this functional was slightly adjusted in 
order to reproduce exactly the experimental binding energy \cite{wapstra03}. 
In this approximation, the experimental charge r.m.s radii 
(compilation of Angeli \cite{angeli04}) 
are reproduced to better than 0.5$\%$. We plotted in 
Fig.\ref{Fig:1} the potentials for the three reactions under investigation 
($^{12}$C+$^{12}$C, $^{12}$C+$^{16}$O, $^{16}$O+$^{16}$O).
Both forces produce the same barrier height and small differences can be observed in 
the barrier's thickness at very low energies. 
\begin{figure}[h]
\center{
\includegraphics[scale=0.51]{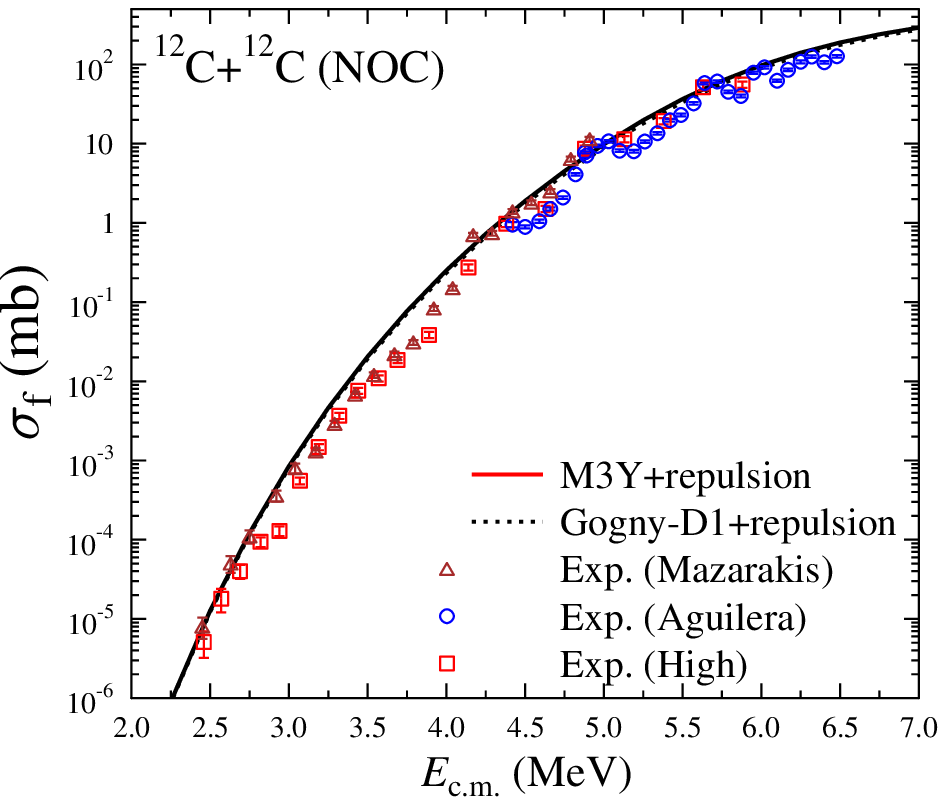}
\includegraphics[scale=0.51]{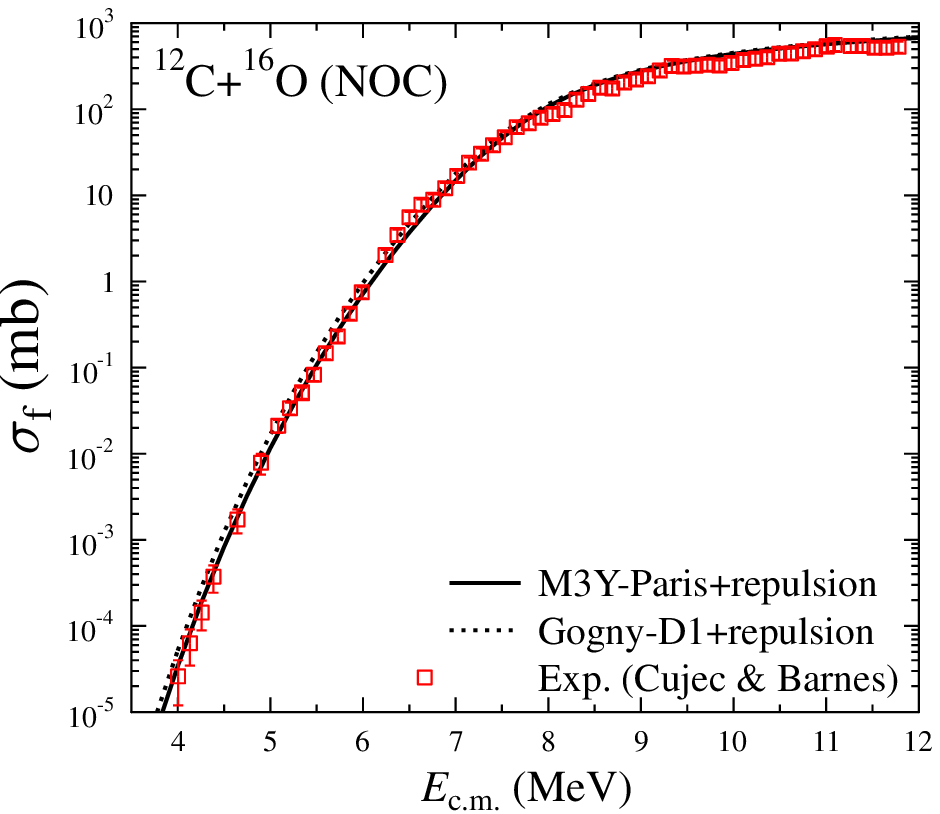}
\includegraphics[scale=0.51]{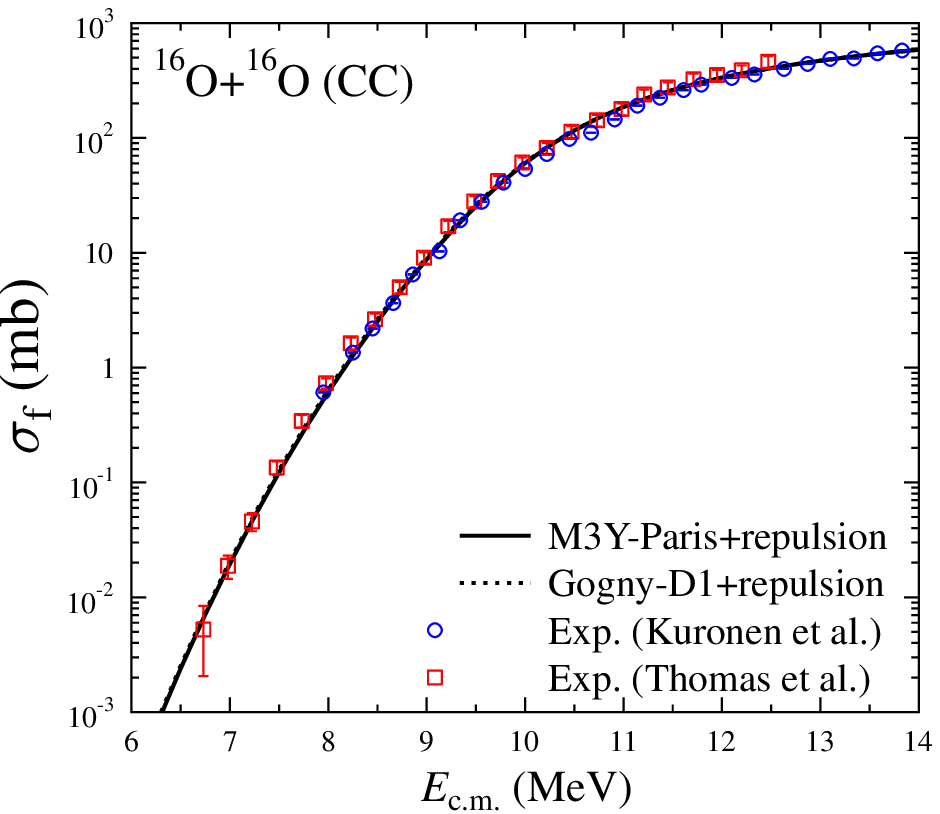}
}
\caption{Fusion cross-sections corresponding to the combinations
and interactions given in Fig.\ref{Fig:1}}
\label{Fig:2}
\end{figure}

The corresponding fusion cross-sections, calculated using the
incoming-wave boundary condition,  are displayed and compared to 
experimental data in Fig.\ref{Fig:2}. The first two reactions 
are well reproduced in the no-coupling (NOC) approximation, whereas
the reaction $^{16}$O+$^{16}$O necessitates the application of 
the coupled-channel method. 
Apart of the $^{12}$C+$^{12}$C cross section which exhibit resonant structures under the 
barrier the data are nicely described. On the other hand we 
could not find any maximum in the $S$-factor, which tells us that for such
light ions the hindrance apparently plays no role.
\begin{figure}[h]
\center{
\includegraphics[scale=0.52]{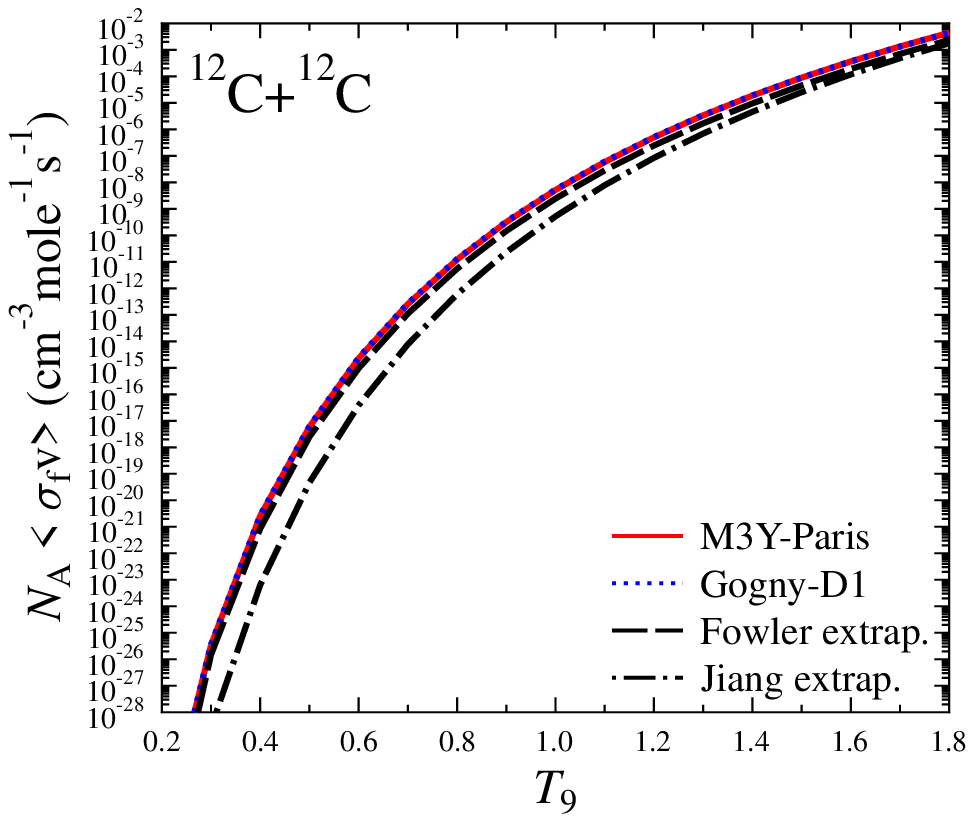}
\includegraphics[scale=0.52]{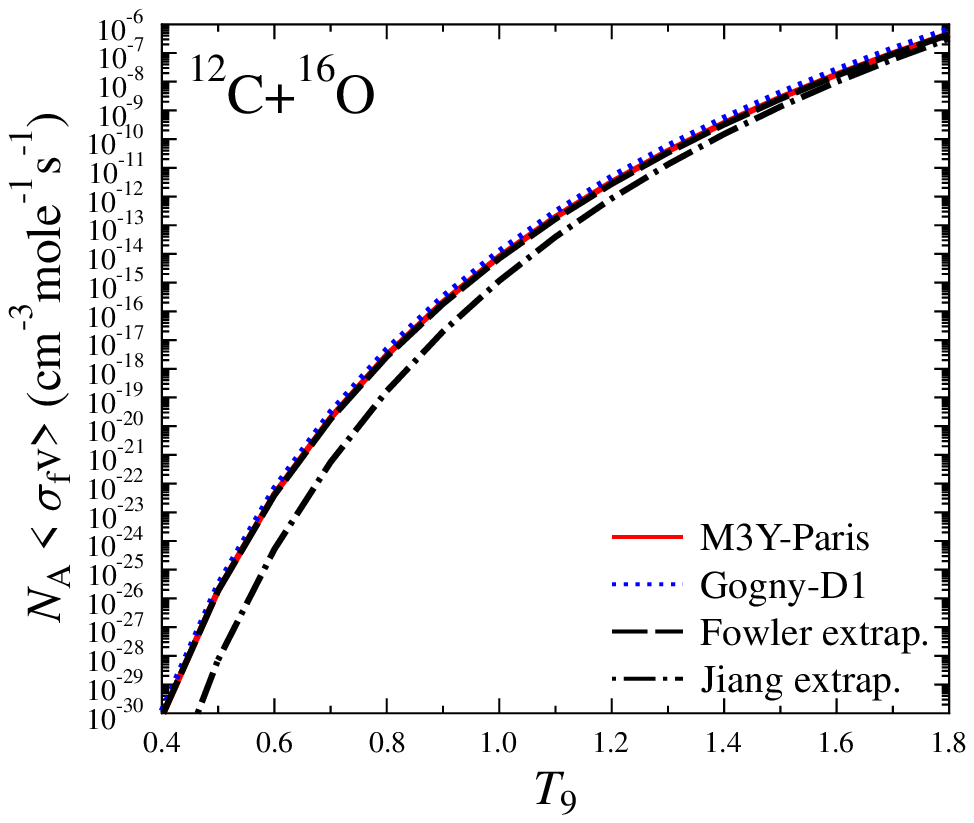}
\includegraphics[scale=0.52]{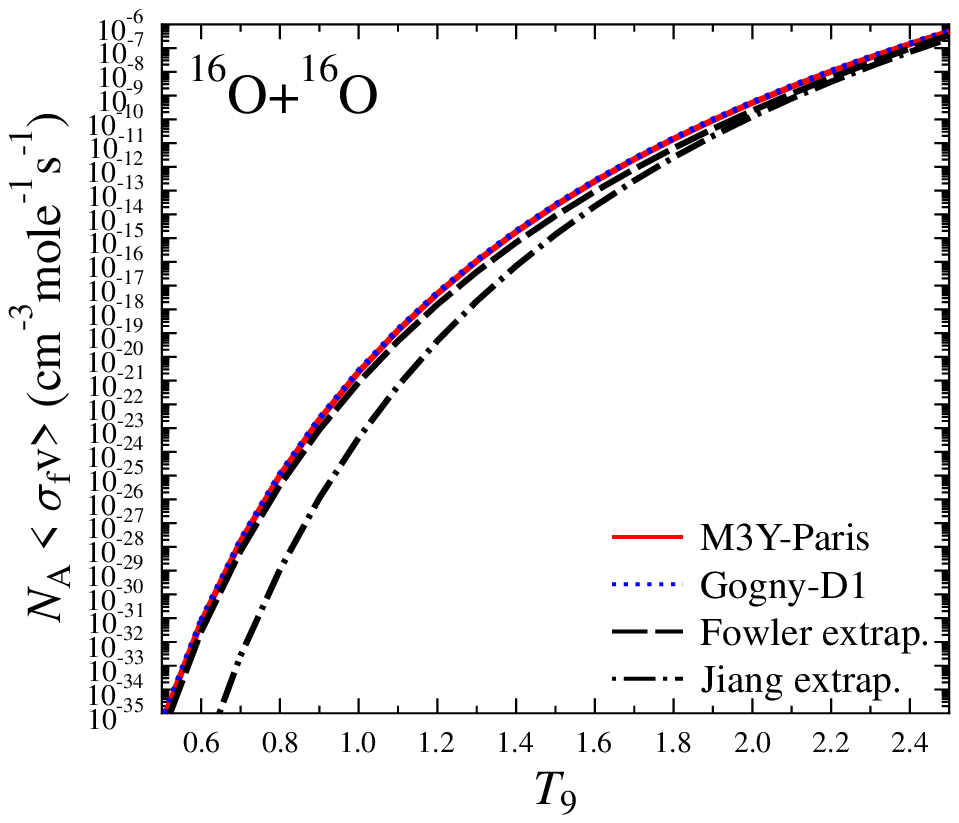}}
\caption{Reaction rates for the three astrophysical reactions. The results 
obtained with the two potentials used in this paper are compared to the
extrapolations of Jiang \cite{jiang07} and Fowler \cite{fow75}}
\label{Fig:3}
\end{figure}

In order to asses the effect of the calculated cross sections at energies
relevant for stellar burning cycles we calculated and present in Fig.\ref{Fig:3}
the reaction rates for the three astrophysical reactions. The fact that our 
predictions are close to the Fowler reaction rate should not be surprising
since no hindrance effect is incorporated in that extrapolation.   
The predicted reaction rate is in all cases only weakly dependent on the type of finite-range
force employed in the calculation of the double-folding potential.
On the other hand in the early phases of the hydrostatic burning stage the 
predicted curve and the Fowler extrapolation are pointing to a speed-up of 
the carbon and oxygen burning compared to the extrapolation formula based on hindrance.

\end{document}